# Circular Dichroism of Chiral Molecules in DNA-Assembled Plasmonic Hotspots

*Luisa M. Kneer[1†], Eva-Maria Roller[1†], Lucas V. Besteiro[2,3], Robert Schreiber[1], Alexander O. Govorov[2] and Tim Liedl[1*]*

[1] Fakultät für Physik and Center for Nanoscience, Ludwig-Maximilians-Universität München, Geschwister-Scholl-Platz 1, 80539 Munich, Germany

[2] Department of Physics and Astronomy, Ohio University, Athens, Ohio 45701, USA

[3] Centre Énergie Matériaux et Télécommunications, Institut National de la Recherche Scientifique, 1650 Boul. Lionel Boulet, Varennes, QC J3X 1S2, Canada

†These authors contributed equally to this work.

**The chiral state of a molecule plays a crucial role in molecular recognition and biochemical reactions. Because of this and owing to the fact that most modern drugs are chiral, the sensitive and reliable detection of the chirality of molecules is of great interest to drug development. The majority of naturally occurring biomolecules exhibit circular dichroism (CD) in the UV-range.[1] Theoretical studies[2,3] and several experiments[4–7] have demonstrated that this UV-CD can be transferred into the plasmonic frequency domain when metal surfaces and chiral biomolecules are in close proximity. Here, we demonstrate that the CD transfer effect can be drastically enhanced by placing chiral molecules, here double-stranded DNA, inside a plasmonic hotspot. By using different particle types (gold, silver, spheres and rods) and by exploiting the versatility of DNA origami we were able to systematically study the impact of varying particle distances on the CD transfer efficiency and to demonstrate CD transfer over the whole optical spectrum down to the near infrared. For this purpose, nanorods were also placed upright on our DNA origami sheets, this way forming strong optical antennas. Theoretical models, demonstrating the intricate relationships between molecular chirality and achiral electric fields, support our experimental findings.**

Circular dichroism (CD) arises when chiral molecules or three-dimensional structures preferentially absorb circularly polarized light of a given handedness. While standard analytical techniques such as UV-VIS absorption, mass spectroscopy, and chromatography are routinely used to identify molecular structures, they usually lack the ability to accurately discern among chiral species (enantiomers). CD spectroscopy is the method of choice for the determination of molecular chirality and for monitoring conformational changes during ongoing chemical reactions or protein folding.[8] One experimental hurdle can be the fact that proteins and other biomolecules absorb mostly in the UV and therefor their CD characterization has to be performed in a frequency range that is hardly accessible due to the strong absorption of water and other buffer solutions below 220 nm.[9] For DNA, e.g., the CD response is studied mainly in the range of 220-300 nm while the absorption reaches considerably further into the UV.[10] Detectable concentrations for DNA are on the order of ~ 25 µg/ml, which corresponds to a concentration of ~ 2 µM for a double strand of 20 base pairs (bp).[11] As already small amounts of enantiomers can have unintended toxicological or pharmacological effects (e.g. thalidomide), it is desired to develop new techniques[12] to increase the limit of detection of the purity of chiral substances with the ultimate goal of single-molecule chirality sensitivity.

It has been shown theoretically[2,13–15] and experimentally[5,7,16,17] that chiral molecules brought into close contact with non-chiral metal nanoparticles can enhance an existing as well as induce an additional CD signal at the plasmon resonance of the nanoparticles, which typically lies in the visible frequency domain for noble metals[3,18–21] This type of chirality transfer is based on the non-zero optical rotatory dispersion (ORD) that biomolecules exhibit in the visible. Implementations of this effect have been realized with biomolecule-nanoparticle heterocomplexes consisting of gold nanoparticles decorated with peptides,[5] cysteine-modified nanoparticles[16], silver nanoparticles synthesized on DNA[17] and polyfluorenes doped with gold nanoparticles.[18] Such nanoscale heterostructues of chiral and achiral components work as surprisingly efficient CD sensors in the visible frequency range. Moreover, the variation of the sizes and shapes of the achiral metal particles allows exploiting a range of beneficial optical properties of plasmonic nanoparticles such as a variety of plasmon resonances and in particular the local field enhancement inside assembled hotspots.[22–24] The versatility of plasmonic behavior can be applied to tune the CD sensor with respect to the desired detection wavelength as well as to amplify even weak chiral signals of low concentrated molecules.[7]

To impart the chirality from chiral molecules onto achiral metal nanoparticles and to take full advantage of the optical properties of nanoparticle hotspots, we here explore well defined and precisely arranged nanoscale heterostructures. We report on the sensing of chiral B-form DNA molecules at visible frequencies with the help of E-field focusing antenna systems consisting of two plasmonic metal nanoparticles arranged with nanometer-precision on DNA origami structures.[25,26] Upon illumination, strong fields arise in the plasmonic hotspots between the particles[24,27] and interact with the chiral molecules placed there. The DNA origami structures fulfill a dual purpose, first as the scaffold arranging the antenna particles and second as being the chiral molecules under investigation. By design, the chiral analyte molecule, *i.e.* the DNA origami structure, is placed exactly inside the plasmonic hotspot formed by the two gold nanoparticles or two gold nanorods.[28] This approach of combining CD transfer with strong electromagnetic field enhancement[29] allows to detect minute amounts of chiral molecules.

Structural DNA nanotechnology enables the assembly of nanoscale three-dimensional structures in a one-pot thermal annealing process. During that reaction, the desired object is obtained by folding a long single-stranded scaffold DNA molecule (here of 7560 nucleotides length) into shape by ~ 200 short synthetic oligonucleotides, so-called staple strands.[25,26] Due to the inherent sequence addressability of the DNA origami structures, it is possible to attach metallic nanoparticles through DNA linkers with nanometer precision and yields up to 99%.[30] We designed several DNA origami structures consisting of sheets of parallel helices with a thickness of one, two or four layers.[31,32] Spherical gold nanoparticles (AuNPs) with 40 nm diameter or gold nanorods (AuNRs) with different aspect ratios (length : diameter of 50 nm : 20 nm = 2.5 and 56 nm : 17 nm = 3.3) were covered with thiolated DNA sequences and attached to the two opposite faces of the DNA structures as depicted in Figure 1a-c (see supporting information S1-S3 for detailed experimental protocols).[33] As a result, the chiral target molecule (here the B-form DNA of the origami) is placed exactly in the hotspot of the nanoantenna (Figure 1d). To reach best field enhancement, we oriented the tips of the AuNRs towards each other as shown in Figure 1c. Note, that the attachment of AuNRs in an upright conformation ("standing" AuNRs) to DNA origami structures is distinct from previously published "lying" orientations.[34,35] By offering 12 linker strands in an area smaller than the cross section of the AuNRs, we took advantage of two effects favoring a standing configuration: i) the crystal facet type of the tips of Au rods are prone to carry higher densities of thiolated DNA linkers[36] and ii) the electrostatic repulsion between the negatively charged

DNA structure and the DNA-covered rods is minimal in a standing orientation (see supporting information S4 for additional details).

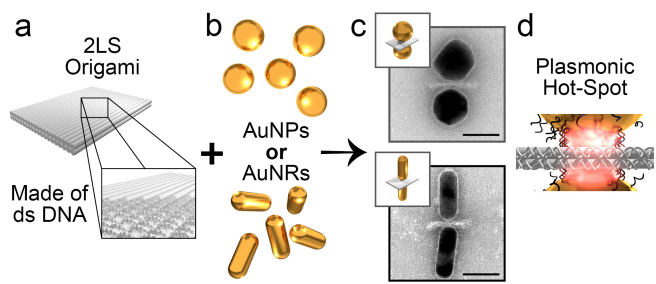

**Figure 1. DNA origami – nanoparticle antennas.** Plasmonic antennas were built by arranging metal nanoparticles on single- or multi-layered DNA origami sheets. **a**, Origami sheets consist of multiple parallel B-form DNA helices. **b**, DNA-functionalized nanoparticles or nanorods were attached via DNA hybridization (DNA on particles not shown). **c**, TEM images of the resulting dimer structures (scale bars: 40 nm). **d**, Design details and DNA attachment strategy. Upon illumination with light strong electric fields arise in the hotspot region between the two particles.

We first performed CD spectroscopy at UV and at visible wavelengths on 2-layer sheets (2LSs) at a concentration of 1 nM. The recorded spectra in Figure 2a exhibits a weak and multimodal CD signal in the UV range arising from the B-form DNA making up the DNA origami structures. Note the pronounced dip-peak shape around the main absorption resonance of DNA molecules at 260 nm (cf. absorption spectra in SI Figure S5a). If the same 2-layer sheets carry two 40 nm AuNPs as described in Fig. 1, we record ~ 30 times stronger CD signals with a corresponding dip-peak shape around the plasmon resonance frequency of the coupled particles at 535 nm. We associate this CD mode to the CD transfer signal from the chiral DNA molecules placed in the plasmonic hotspot as described in theory by Govorov *et al.*.[2,15,37] If the optical rotatory dispersion from a chiral molecule has a non-zero signal in the visible frequency range, which overlaps with the absorption of the plasmonic particles, the CD signal can be transferred and detected at the plasmon resonance frequency. In this case, the absorption process contributing to the CD signal takes place inside a non-chiral plasmonic dimer, which interacts with a chiral dipole of a biomolecule. For single spherical NPs, the molecule-plasmon interaction is mostly dipole-like,[5] whereas, for the dimers with a hotspot, it becomes highly multipolar.[37] Importantly, the high electromagnetic field enhancement in the plasmonic hotspots allows us to detect the transferred CD signal at much lower analyte concentrations compared to what is needed to obtain discernible signals in the UV-CD mode. This effect is shown in Figure 2a and occurs across the entire optical frequency window. To demonstrate the CD transfer into the near infrared we assembled two differently sized AuNR antennas. The red graph shows the CD pick up of an antenna system with two AuNRs with an aspect ratio of 2.5 that exhibits a coupled longitudinal resonance at 700 nm. By placing the AuNRs upright as described above and in the supplementary information, we obtain a 140-fold stronger signal compared to the UV CD. For AuNRs with an aspect ratio of 3.3 the longitudinal resonance shifts to 810 nm (see dark red graph in Figure 2a) and a 300-fold enhancement is achieved. We thus conclude that the increased field intensity inside the hotspots permits to sense the chirality of molecules at concentrations way below the UV detection limit. (The corresponding absorption spectra as well as more TEM images can be found in SI Figure S5a and SI Note S6.)

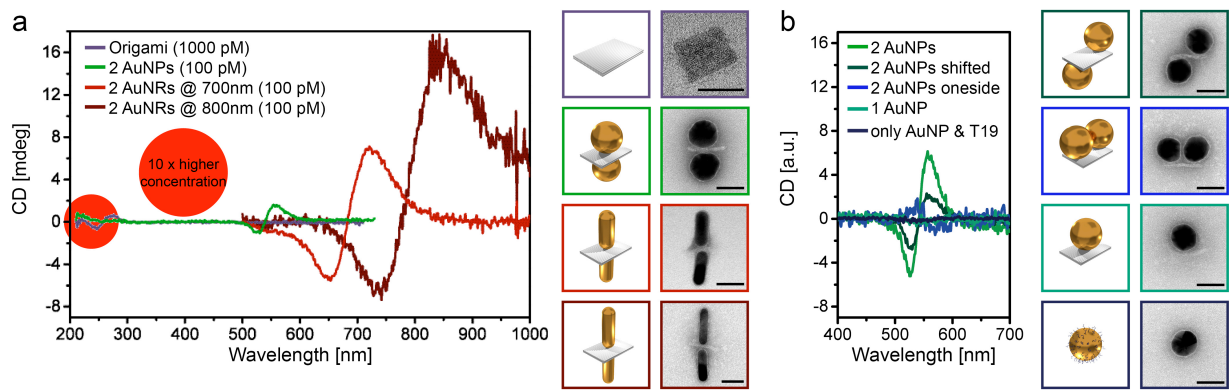

**Figure 2. CD measurements of bulk nanoantenna in solution.** The UV-CD of B-form DNA is transferred into the plasmonic frequency range of the nanoantennas. **a**, The characteristic shape of the UV-CD signal of a 2-layered DNA origami sheet is discernable at a concentration of 1 nM. The signal is transferred to the red and amplified by a factor of 30 for AuNPs antenna structures (green) and up to 300-fold for AuNRs antenna structures (red and dark red). **b**, By comparing the AuNP dimer (green) with control structures at equal concentration plasmonic CD can be excluded (cf. text). All illustrations and TEM images are shown with 40 nm scale bars.

In order to underline the antenna power as well as to exclude plasmonic chirality effects – i.e. the possibility of two not fully spherical particles accidently forming a chiral object – we performed a set of control experiments (Figure 2b). First, the 2LS AuNP dimer was altered by shifting both AuNPs away from the center of the origami faces towards their edges. This mainly decreases the field intensity in the hotspot as the distance between the particle surfaces increases from ~ 7.8 nm (Fig. S7) (AuNPs opposite of each other, green frame in Fig. 2a) to ~ 12.6 nm (AuNPs shifted towards each other, dark green frame in Fig. 2b). Indeed, the recorded CD intensity of the shifted dimer drops to less than half of the value of the opposing dimer. When both gold nanoparticles are placed on the same side of the 2LS origami, the DNA double strands of the origami structure are no longer in the plasmonic hotspot (blue frame in Fig. 2b). As a result, we obtained a barely measurable signal exhibiting a peak-dip shape (blue curve in Fig. 2b). We attribute this CD signal to the unbound single-stranded DNA linkers that are covering the gold particles and some of which are present in the ~ 3.5 nm wide gap between the particles. The fact that we consistently observe strong CD signals with dip-peak shape in the antenna configuration with double-stranded DNA (dsDNA) inside the hotspot and we only observe a very weak signal in a configuration with no dsDNA inside the hotspot but with even stronger particle-particle coupling (3.5 nm vs. 7.8 nm surface-to-surface distance) excludes the possibility of unintended formation of chiral plasmonic structures. Note, that all measurements were performed in solution at equal concentrations (cf. absorption spectra in SI Figure S5b) and that we are thus recording an ensemble average. In experiments with only one AuNP attached to the origami structures (cyan curve and frame in Fig. 2b) or with the individual linker-functionalized AuNPs (marine curve and frame) no CD occurs. We conclude, that placing chiral molecules of interest in a plasmonic hotspot conformation rather than just close to a single plasmonic particle is crucial for high detection sensitivity. This underlines that due the combination of two effects - the CD transfer and the plasmonic enhancement – it is possible to create a very sensitive detection device for chiral analytes at visible wavelengths.

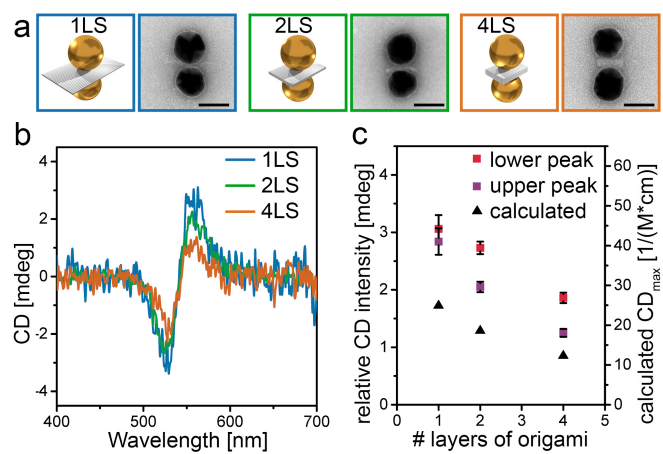

**Figure 3. CD spectra of origami dimer structures with a varying number of DNA layers. a,** Schematic illustrations and corresponding TEM images of a 1-layer sheet, 2-layer sheet and 4-layer-sheet (scale bars: 40 nm). **b,** CD spectra of the three structures reveal decreasing CD intensity with increasing gap size and thus decreasing hotspot strength. The spectra are normalized with respect to the concentrations of the assembled dimes. **c,** The trend for the maximum CD peak intensities of the upper and lower wavelength peak from the experimental data matches that of the calculated maximum CD intensities. The error bars reflect the error of the concentrations of assembled dimers.

The influence of the plasmonic hotspot intensity on the resulting CD signal strength was further studied by comparing the CD spectra of 40 nm AuNP dimer structures with varying numbers of DNA layers in the gap. For this purpose, three different DNA origami structures were assembled: a one-layer sheet (1LS), the 2LS and a four-layer sheet (4LS) (Figure 3a). The gap size between the two attached AuNPs theoretically varies between 5 nm for the 1LS (2 nm thickness for one layer of dsDNA + 2 x 1.5 nm linker DNA), 7.5 nm (2 x 2 nm thickness for one layer of dsDNA + 0.5 nm gap between the two layers + 2 x 1.5 nm linker DNA) to 12.5 nm (4 x 2 nm thickness for one layer of dsDNA + 3 x 0.5 nm gap between the two layers + 2 x 1.5 nm linker DNA). The measured gap sizes are 5.0 nm, 7.8 nm and 12.4 nm respectively (cf. SI Note S7). The resulting CD spectra of the three different structures are shown at equal concentration (cf. absorption spectra SI Figure S5c) in Figure 3b. Increasing the gap size by increasing the amount of DNA layers between the particles results in a decrease of the CD signal peak intensities as depicted in Figure 3c. This trend, however, is not as drastic as one could naively expect. In fact, also the amount of the analyte molecules – the double-stranded DNA within the hotspot – grows with growing gap size. This partly counterbalances the effect of the decreasing hotspot intensity with increasing particle distances. Yet, our results are in very good agreement with the calculated maximal CD intensity of a 40 nm AuNP dimer (black triangles in Figure 3c). For the numerical simulations, we assumed the nominal gap sizes of 5 nm, 7.5 nm and 12.5 nm and placed one chiral dipole per layer of DNA and an additional 2 dipoles to account for the double strands formed by the handle sequences with the linker DNA of the particles on both sides. This resulted in 3, 4 and 6 dipoles respectively. We applied the model and computational formalism described in Zhang *et al.*.[37] The simulated peak intensities match those of all our experiments, which supports our theoretical understanding and highlights the level of control offered by DNA origami-assembled nanoantennas.[38]

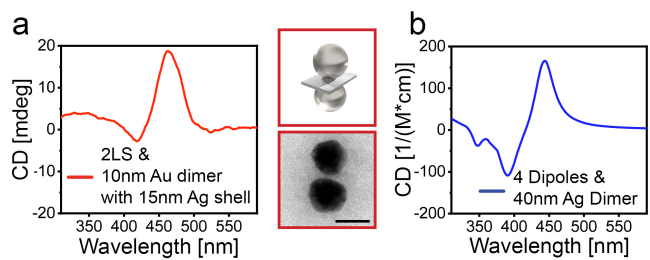

**Figure 4. CD spectra of Ag shell antenna. a,** The CD signal of 2-layer sheets origami carrying two 10 nm AuNPs overgrown with a 15 nm Ag shell (red) and **b,** the simulated spectrum of a dimer with 40 nm Ag spheres and 4 chiral dipoles in a 5 nm gap (blue) are in strikingly good agreement (scale bar: 40 nm).

Finally, we showed that the use of other materials for the antenna elements enables us to shift the CD signal into yet another wavelength range. Two 10 nm AuNPs placed on the 2LS were overgrown with a 15 nm thick shell of silver (see supporting information S8 for experimental protocol). Due to the absorption resonance of these silver shell particles at ~ 430 nm also the CD signal occurred in this wavelength range as shown in red in Figure 4. Under the assumption that the relatively thick silver shell dominates the plasmonic behavior in this system, we performed numerical simulations with a pure 40 nm Ag dimer that hosts 4 dipoles in its 5 nm gap (Figure 4). Again, we chose 4 dipoles to account for the two origami layers of dsDNA and the double strands formed by the handle sequences with the linker DNA of the two particles. Strikingly, the theoretical spectrum based on the theory from Zhang *et al.*[37] reproduces both the shape of the experimental data centered at the silver plasmon resonance and the strong signal amplification. As mentioned above, the plasmonic CD structure is not just a single plasmonic peak, like it was initially observed for single AuNPs[5], but a complex structure arising from multipolar plasmonic resonances in our hybrid plasmonic system.[37] This character and related mechanism of plasmon induced CD is typical for systems with plasmonic hotspots.[37,39]

In summary, we demonstrated that our nanoantenna heterostructures facilitate the observation of induced chiral transfer of molecular CD signals from the UV to the visible and infrared frequency range at analyte concentrations below 100 pM. Control experiments with carefully designed DNA structures allow us to unequivocally link the occurrence of the CD signals at the plasmonic resonances to a transfer process from the primary CD signal of the chiral molecules located within the plasmonic hotspot. Numerical simulations match our quantitative analysis and support our understanding of the strong signal amplifications in the plasmon-enhanced electric fields. The inherent sequence- and consequently site-addressability of DNA origami structures will allow the construction of antenna templates for various chiral analytes precisely placed in the plasmonic hotspot. Future experiments could aim at depositing such addressable antennas on solid substrates[5,40] with the ultimate goal of single-molecular chirality detection.[41]

**Supporting Information is available online.**

**Corresponding Author**

*Correspondence and request for materials should be addressed to: tim.liedl@lmu.de

**Author Contributions**

TL and RS conceived the research, EMR and LK performed the experiments. AOG and LVB contributed the simulations. LK, EMR and TL wrote the manuscript.

The authors declare no competing financial interests.

ACKNOWLEDGMENT

This work was supported by the Volkswagen Foundation, the DFG through the Nanosystems Initiative Munich (NIM), the European Commission under the Seventh Framework Programme (FP7) as part of the Marie Curie Initial Training Network, EScoDNA (GA no. 317110) and the ERC grant ORCA, agreement n° 336440.